\documentclass{iopart}

\usepackage{epsfig}
\usepackage{amssymb}

\begin{document}

\title[Symmetries within EDF methods]{Breaking and restoring symmetries within the nuclear energy density functional method}

\author{T. Duguet$^{1,2}$ and J. Sadoudi$^1$}

\address{$^1$ CEA, Centre de Saclay, IRFU/Service de Physique Nucl{\'e}aire, F-91191 Gif-sur-Yvette, France}
\address{$^2$ National Superconducting Cyclotron Laboratory and Department of Physics and Astronomy,
Michigan State University, East Lansing, MI 48824, USA}
\eads{\mailto{thomas.duguet@cea.fr}, \mailto{jeremy.sadoudi@cea.fr}}

\begin{abstract}
We review the notion of symmetry breaking and restoration within the frame of nuclear energy density functional methods. We focus on key differences between wave-function- and energy-functional-based methods. In particular, we point to difficulties to formulate the restoration of symmetries within the energy functional framework. The problems tackled recently in connection with particle-number restoration serve as a baseline to the present discussion. Reaching out to angular-momentum restoration, we identify an exact mathematical property of the energy density $E^{LM}(\vec{R})$ that could be used to constrain energy density functional kernels. Consequently, we suggest possible routes towards a better formulation of symmetry restorations within energy density functional methods.
\end{abstract}


\section{Introduction}

\subsection{Spontaneous symmetry breaking}

Symmetries are essential features of classical and quantal systems. For the latter in particular, symmetries characterize the energetics of the system and provide transition matrix elements of operators with specific selection rules. In nuclear systems for example, electromagnetic and electro-weak decays display patterns associated with such selection rules.

On the other hand, certain emergent phenomena relate to the spontaneous breaking of those symmetries~\cite{yannouleas07a}. In the thermodynamic limit, i.e. when the number of particles $N$ and the volume $V$ of the system go to infinity such that $N/V$ remains constant, a state with lower symmetry than the Hamiltonian can be rigorously used as an effective ground-state of the system. Such a state is a linear superposition of nearly-degenerate eigenstates, i.e. it is a wave-packet. In finite systems however, quantum fluctuations make such a wave-packet to relax into the symmetry-conserving ground-state and cannot be ignored; i.e. the concept of spontaneous symmetry breaking is only an intermediate description of the system that arises within certain approximations and symmetries must eventually be restored. Still, it makes physical sense to go through such an intermediate description as pseudo spontaneously-broken symmetries (i) relate to specific features of the inter-particle interactions, (ii) characterize internal correlations and (ii) leave clear fingerprints in the observed excitation spectrum of the system.

\begin{table}[hb]
\caption{\label{symmetrybreaking} Links between the spontaneous breaking of translational, rotational and particle-number symmetries and features of the nuclear force, correlations in the internal motion of nucleons and patterns in the excitation spectrum.}
\small\rm
\begin{tabular}{|c|c|c|c|}
\hline
Invariance & $V^{NN}$ & Internal correlations &  Excitation patterns \\
\hline
Spatial translation & Short range & Spatial localization & Surface vibrations \\
Gauge rotation & $S$-wave attraction  & Pairing  &  Energy gap \\
Spatial rotation & Quad.-quad. component & Angular localization & Rotational bands  \\
\hline
\end{tabular}
\end{table}

In atomic nuclei, several symmetries, if allowed, tend to break spontaneously in approximate descriptions based on the mean-field concept. The most important ones relate to the invariance of the nuclear Hamiltonian $H$ under spatial translations and rotations as well as to the gauge invariance associated with particle-number symmetry. As described in Tab.~\ref{symmetrybreaking}, the spontaneous breaking of these three symmetries relates to the short-range and dominant quadrupole-quadrupole terms of the nucleon-nucleon interaction as well as to its strong attraction in the $L=0$ partial-wave of relative motion. In particular, the strong attraction in the $L=0$ partial-wave in particular generates a $S$-wave di-neutron (di-proton) virtual state at almost zero scattering energy that is the precursor of neutron (proton) Cooper pairs and superfluidity in the nuclear medium. Even though such symmetries must be eventually enforced, their underlying breaking impacts the low-lying spectroscopy of finite nuclei through the presence of surface vibrational excitations, rotational bands and a gap in the individual excitations of even-even nuclei, respectively~\cite{ring80a}. Parity and time-reversal are other good symmetries of $H$ that can be spontaneously broken, while isospin symmetry is only approximate in the first place.

\subsection{Wave-function-based methods}

\begin{center}
\begin{figure}
\hspace{3.5cm} \includegraphics[width=6.cm, clip]{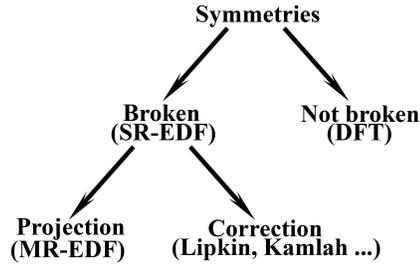}
\caption{(Color online) Schematic representation of the different strategies followed by many-body methods regarding the treatment of symmetries, e.g. in Density Functional Theory and nuclear Energy Density Functional approaches.}
\label{fig:dftvsedf}
\end{figure}
\end{center}

\begin{figure}
\hspace{2.5cm} \includegraphics[width = 0.53\textwidth, keepaspectratio]{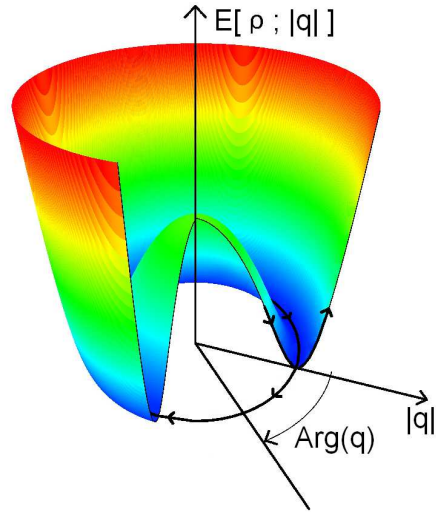}
\caption{(Color online) Schematic view of the energy as a function of the phase and magnitude of the order parameter $q$ of a spontaneously broken symmetry.}
\label{fig:hat}
\end{figure}

\begin{figure}
\hspace{0.5cm} \includegraphics[width = 0.45\textwidth, keepaspectratio]{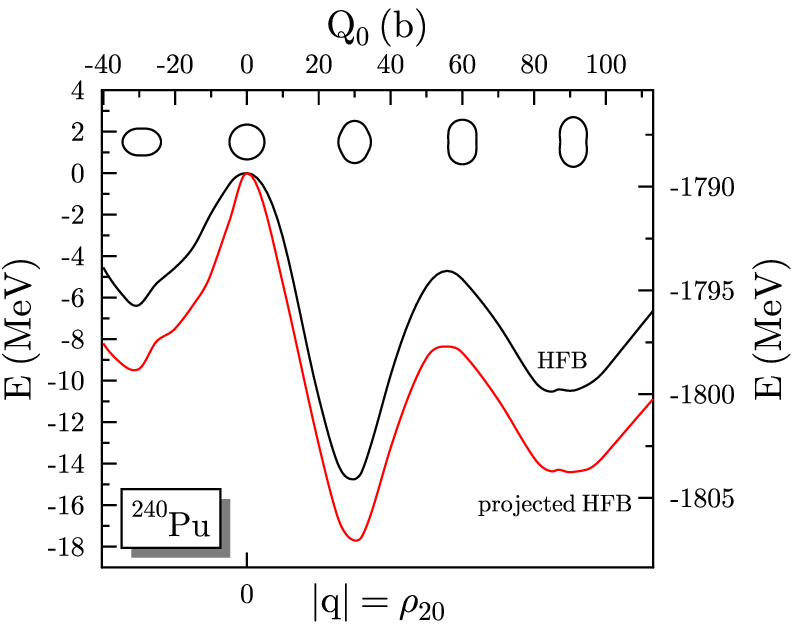} \hspace{0.8cm} \includegraphics[width = 0.45\textwidth, keepaspectratio]{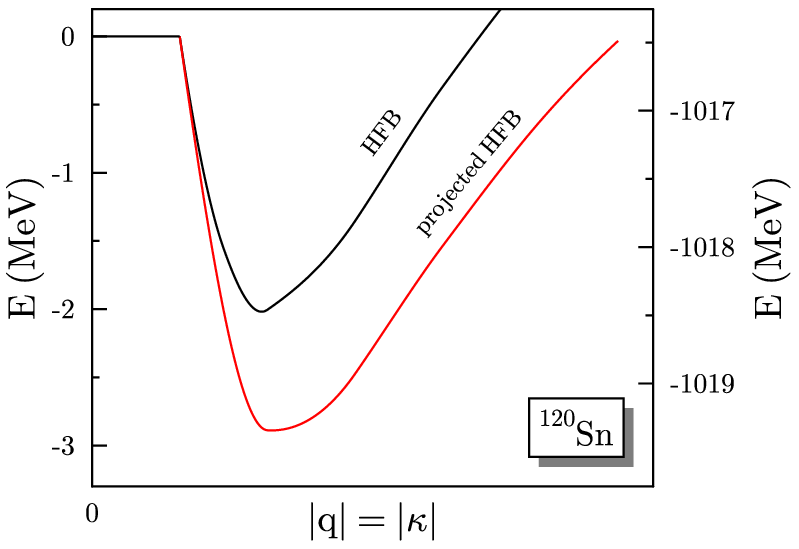}
\caption{(Color online) Energy gain from spontaneous symmetry breaking and symmetry restoration as a function of the magnitude of the order parameter $q$. Left: breaking and restoration of rotational symmetry in the ground state of $^{240}$Pu as a function of the axial quadrupole moment of the single-nucleon density distribution (adapted from Ref.~\cite{bender04a}). Right: breaking and restoration of neutron-number symmetry in the ground state of $^{120}$Sn as a function of the norm of the anomalous pair density (adapted from Ref.~\cite{bender07a}).}
\label{fig:examplemredf}
\end{figure}

As schematically shown in Fig.~\ref{fig:dftvsedf}, quantum many-body methods separate into two categories as for the way symmetries are dealt with, i.e. (i) methods enforcing symmetries throughout and (ii) those that explicitly single out the intermediate breaking of symmetries. Although hybrid approaches that allow the breaking of some symmetries while enforcing the others can be set up, the present work focuses on those that strongly rely on the concept of symmetry breaking, i.e. methods whose philosophy, apart for computational constraints, is to allow all symmetries to break spontaneously a priori. The breaking of each symmetry is monitored by the magnitude and the phase of an order parameter $q$, such that the (approximate) energy is independent of the phase as schematically shown in Fig.~\ref{fig:hat}. This corresponds to the fact that a spontaneous symmetry breaking is accompanied by the presence of a zero-energy Goldstone mode. Of course, that a certain symmetry does break spontaneously usually depends on the number of elementary constituents of the system under consideration. For example, while translational symmetry (strongly) breaks in all nuclei, particle-number symmetry tend to (weakly) break in all but doubly-magic nuclei whereas rotational symmetry remains unbroken if either the neutron number or the proton number is "magic"\footnote{The fact that the neutron or proton number is magic is not known a priori but is based on a posteriori observations and experimental facts. In particular, the fact that traditional magic numbers, i.e. $N,Z=2, 8, 20, 28, 50, 82, 126$, remain as one goes to very isospin-asymmetric nuclei is the subject of intense on-going experimental and theoretical investigations~\cite{sorlin08a}.}. Figure~\ref{fig:examplemredf} displays the correlation energy incorporated in $^{240}$Pu and $^{120}$Sn ground-states energy through the spontaneous breaking of rotational and particle-number symmetries, respectively. Such symmetry breakings may account for up to $20$ MeV correlation energy out of about $2$ GeV binding energy, i.e. for about $2\%$, which is much larger than the targeted accuracy on nuclear masses. Incorporating such correlation energies through symmetry-conserving approaches, e.g. configuration interaction methods, would necessitate tremendous computational efforts in such heavy open-shell nuclei.

As already stated, methods authorizing the breaking of symmetries at a certain level of approximation must eventually restore them in a second stage. In wave-functions based methods, the symmetry breaking step, e.g. the symmetry unrestricted Hartree-Fock-Bogoliubov approximation, relies on minimizing the average value of the Hamiltonian for a trial wave-function that does not carry good quantum numbers, i.e. which mixes irreducible representations of the symmetry group of interest. Restoring symmetries amounts to using an enriched trial wave-function that does carry good quantum numbers. In terms of the schematic "mexican-hat" picture of Fig.~\ref{fig:hat}, this corresponds to incorporating zero-energy fluctuations associated with the phase of the order parameter\footnote{Although it is not the focus of the present work, the restoration of symmetries must be complemented with the inclusion of collective quantum correlations associated with the fluctuations of the {\it magnitude} of the order parameter. This can be done through the so-called Generator Coordinate Method (or its energy density functional counterpart) that is formally identical to projection methods discussed here.}. One typical approach is to project out from the symmetry-breaking trial state the component that belongs to the intended irreducible representation~\cite{ring80a}. Figure~\ref{fig:examplemredf} shows that doing so for rotational and particle-number symmetries adds a few MeV correlation energy to the ground-state binding energy of heavy nuclei. This is still significant compared to the few hundreds keV targeted accuracy on nuclear masses. As shown in Fig.~\ref{fig:dftvsedf}, a variant consists of performing the projection only approximately such that the calculation boils down to the minimization of a {\it corrected} energy expressed in terms the original symmetry-breaking state. Typical examples are Lipkin~\cite{lipkin60a,doba09a} of Kamlah approximate projection methods~\cite{valor96a,kamlah68a}. While it is likely that the strongly broken translational symmetry can be safely treated through such approximate projection methods\footnote{Such a statement is to be taken with a grain of salt for rather light nuclei~\cite{rodriguezguzman04a}.}, it is still unclear whether the same is true for weakly broken symmetries such as particle number symmetry or rotational symmetry in transitional nuclei.

\subsection{The nuclear energy density functional method}

Wave-function-based projection methods and their variants are well formulated quantum mechanically~\cite{ring80a}. The goal of the present paper is to discuss their Energy Density Functional (EDF) counterparts~\cite{bender03a} which have been {\it empirically adapted} from the former to deal quantitatively with properties of nuclei. The nuclear EDF method does rely heavily on the concept of spontaneous symmetry breaking and (approximate) restoration. In that sense, it is intrinsically a two-step approach. The so-called single-reference EDF (SR-EDF) method has been originally adapted from the symmetry-unrestricted Hartree-Fock-Bogoliubov method by using a {\it density-dependent} effective Hamilton "operator"~\cite{negele72a}. Later on, the approximate energy was formulated directly as a possibly richer functional of one-body density matrices computed from a symmetry-breaking product-state of reference. The second step, carried out through the multi-reference (MR) extension of the SR-EDF has been adapted from the projected Hartree-Fock-Bogoliubov method. Such a step necessitates a prescription to extend the SR energy functional\footnote{The density-dependence of the effective Hamilton operator in more standard formulations.} associated to a single auxiliary state of reference, i.e. a diagonal energy kernel, to the non-diagonal energy kernel associated with a pair of reference states (see Sec.~\ref{MREDFmethod}). Constraints based on physical requirements have been worked out that limit the number of possible prescriptions to do so~\cite{robledo07a}. Still, pathologies~\cite{almehed01a,anguiano01a,doba07a} of MR-EDF calculations have been recently identified and corresponding cures~\cite{lacroix09a,bender09a,duguet09a} have been proposed. Besides the actual successes of nuclear EDF calculations~\cite{bender03a}, the work of Refs.~\cite{lacroix09a,bender09a,duguet09a,lacroix09b} demonstrates that nuclear SR- and MR-EDF methods must be further constrained to become satisfactory many-body approaches to finite Fermi systems. The first goal of the present work will be to reformulate, focusing on group-theory considerations, concerns about MR-EDF calculations that have been dealt with in Refs.~\cite{lacroix09a,bender09a,duguet09a,lacroix09b}. Our second objective will be to provide a new mathematical property that could be used in the case of angular-momentum restoration to constrain the form of basic EDF kernels at play.

Given the efforts needed to better formulate EDF methods, one may question the necessity to stick to such approaches rather than to use a wave-function-based approach that strictly computes the energy from a (state-/density-independent) Hamiltonian, e.g. through many-body perturbation theory. Although the breaking (up to a few tens of MeV) and the restoration (up to a few MeV) of symmetries bring in both types of methods essential correlations that vary rapidly with nucleon numbers, incorporating the bulk of correlations\footnote{We take as a loose definition of bulk correlations the correlation energy computed beyond a genuine Hartree-Fock approximation in terms of the vacuum (low-momentum~\cite{Bogner:2003wn}) nuclear Hamiltonian for nuclei that do not break any symmetry besides translational invariance, i.e. doubly-magic nuclei.}  (hundreds of MeV) requires involved wave-function-based calculations~\cite{Roth:2005ah} that are still impractical for heavy open-shell nuclei. The power of the EDF approach is to parameterize bulk correlations under the form of a functional of the one-body density (matrices) such that systematic calculations of heavy nuclei are tractable. The success of the overall approach, based on the resummation of bulk correlations into the EDF kernel and the further breaking and restoration of symmetries, relies on the empirical decoupling of the different categories of correlations at play, i.e. on the different scales that characterize them (see Tab.~\ref{scalescorrelations}), and on the fact that quickly varying correlations with the filling of nuclear shells are explicitly accounted for through symmetry breaking and restoration\footnote{Here we have in mind to add the fluctuations of the magnitude of the order parameter.}.

\begin{table}[hb]
\caption{\label{scalescorrelations} Schematic classification of correlation energies as they naturally appear in nuclear EDF methods. The quantity $A_{{\rm val}}$ denotes the number of valence nucleons while $G_{{\rm deg}}$ characterizes the degeneracy of the valence major shell.}
\small\rm
\begin{tabular}{|l|l|l|l|}
\hline
Correlation energy & Treatment & Scale &  Vary with \\
\hline
Bulk & Summed into EDF kernel & $\sim 8\, A$ MeV  & $A$ \\
Static collective & Finite order parameter $q$  & $\lesssim 25$ MeV  & $A_{{\rm val}}, G_{{\rm deg}}$  \\
Dynamical collective & Fluctuations of $q$ & $\lesssim 5$ MeV &   $A_{{\rm val}}, G_{{\rm deg}}$ \\
\hline
\end{tabular}
\end{table}

\subsection{Density functional theory}

As the aim of the present paper is to raise questions about the treatment of symmetries within the nuclear EDF method, let us make a few relevant statements about Density Functional Theory (DFT)
\cite{hohenberg64,kohn64a,lecturenotesFNM} that provides a formal framework
to obtain the ground-state energy and one-body density of electronic many-body systems. It has become customary in nuclear physics to assimilate the SR-EDF method, eventually including corrections {\it a la} Lipkin or Kamlah, with DFT, i.e. to state that the Hohenberg-Kohn theorem underlays nuclear SR-EDF calculations~\cite{bei74a,fayans,lecturenotesLGV,bulgac3,furnstahl05a}. This is a misconception as distinct strategies actually support both methods. Whereas the SR-EDF method minimizes the energy with respect to a symmetry-breaking trial density(ies), DFT relies on an energy functional whose minimum must be reached for a one-body density that possesses {\it all} symmetries of the actual ground-state density, i.e. that displays fingerprints of the symmetry quantum numbers of the underlying exact ground-state~\cite{fertig00a}. As a matter of fact, generating a symmetry-breaking solution is problematic in DFT, as it lies outside the frame of the Hohenberg-Kohn theorem, and is usually referred to as the {\it symmetry dilemma}. To by-pass the symmetry dilemma and grasp kinematical correlations associated with good symmetries, several reformulations of DFT have been proposed over the years, e.g. see Refs.~\cite{gross88a,gorling93a}, some of which are actually close in spirit to the nuclear MR-EDF method~\cite{gross88a}.

Recent efforts within the nuclear community have been devoted to formulating a Hohenberg-Kohn-like theorem in terms of the internal density, i.e. the matter distribution relative to the center of mass of the self-bound system~\cite{engel07a,messud09a}. Together with an appropriate Kohn-Sham scheme~\cite{messud09a}, it allows one to reinterpret the SR-EDF method as a functional of the internal density rather than as a functional of a translational-symmetry-breaking density. This constitutes an interesting route whose ultimate consequence would be to remove entirely the notion of breaking and restoration of symmetries from the EDF approach and make the SR formulation a complete many-body method, at least in principle. To reach such a point though, the work of Refs.~\cite{engel07a,messud09a} must be extended, at least, to rotational and particle-number symmetries, knowing that translational symmetry was somewhat the easy case to deal with given the explicit decoupling of internal and center of mass motions.

\subsection{Outline}

The paper is organized as follows. Section~\ref{symmetryingredients} provides the minimal set of group theory considerations that are needed to go through the remaining of the paper. Section~\ref{wfbasedmethod} discusses symmetry breaking and restoration within a wave-function based method while Sec.~\ref{EDFmethod} provides the analog within the EDF context. In Sec.~\ref{EDFmethod}, we focus on difficulties encountered to formulate symmetry restorations within the energy functional approach. Finally, Sec.~\ref{newconstraints} briefly discusses a mathematical property associated with angular-momentum conservation that could used to formulate new constraints on off-diagonal energy functional kernels at play in MR-EDF calculations.

\section{Symmetry group}
\label{symmetryingredients}

Let us consider an arbitrary continuous compact group ${\cal G}\equiv\{{\cal R}(g)\}$ parameterized by $r$ real parameters $g\equiv\{g_i ; i=1,\ldots, r\}$ and whose transformations leave $H$ invariant. We denote by $v_{{\cal G}}$ the volume of the group
\begin{equation}
v_{{\cal G}} \equiv \int_{{\cal G}} dm(g) \,\,\, ,
\end{equation}
where $m(g)$ is the invariant measure on ${\cal G}$. Having in mind to deal more specifically with particle number and rotational symmetries, we further consider ${\cal G}$ to be a Lie group, although this is not mandatory. We thus introduce the set of infinitesimal generators ${\cal C}=\{{\cal C}_i ; i=1,\ldots, r\}$ that make up the Lie algebra and in terms of which any transformation of the group can be expressed. The Casimir of the group built from the infinitesimal generators and a non-degenerate invariant bilinear form is denoted by $\Lambda$. We also denote by $R(g)$ ($C$) a unitary representation of ${\cal R}(g)$ (${\cal C}$) on the Fock space of quantum mechanics and by $S^{\lambda}_{ab}(g) \equiv \langle \Theta^{\lambda a} | \, R(g) \, | \Theta^{\lambda b} \rangle$ the matrix elements of the unitary irreducible representation labeled by $\lambda$. Noticing that $S^{\lambda}_{ab}(0)=\delta_{ab}$ for all $\lambda$, the action of two successive
transformations and the unitarity of the representation can be both read off the following identity
\begin{eqnarray}
\sum_{c} S^{\lambda \, \ast}_{ca}(g') \, S^{\lambda}_{cb}(g)  &=& \sum_{c} S^{\lambda}_{ac}(-g') \, S^{\lambda}_{cb}(g) = S^{\lambda}_{ab}(g\!-\!g') \,\,\, ,
\label{unitarity}
\end{eqnarray}
where $-g$ and $g\!-\!g'$ symbolically denote the parameters of transformations $R^{-1}(g)$ and $R^{-1}(g')R(g)$, respectively. States $| \Theta^{\lambda a} \rangle$ span the irreducible representation $\lambda$ whose degree is $d_{\lambda}$. They are eigenstates of the Casimir $\Lambda$ and of a chosen generator $C_{i}$
\begin{eqnarray}
\Lambda \, | \Theta^{\lambda a} \rangle &=&  l(\lambda) \, | \Theta^{\lambda a} \rangle \,\,\, , \\
C_i \, | \Theta^{\lambda a} \rangle &=&  g(a) \, | \Theta^{\lambda a} \rangle \,\,\, ,
\end{eqnarray}
where eigenvalues $l(\lambda)$ and $g(a)$ are functions of labels $\lambda$ and $a$, respectively, with $a$ running over $d_{\lambda}$ values. A so-called {\it irreducible tensor operator} $T^{\lambda}_{a}$ and a state $| \Theta^{\lambda a} \rangle$ transform according to
\begin{eqnarray}
R(g) \, T^{\lambda}_{a} \, R(g)^{-1}  &=& \sum_{b}  T^{\lambda}_{b} \, S^{\lambda}_{ba}(g) \,\,\, ,\label{eq:ten:def1} \\
R(g) \, | \Theta^{\lambda a} \rangle &=& \sum_{b}  | \Theta^{\lambda b} \rangle \, S^{\lambda}_{ba}(g) \,\,\, . \label{eq:ten:def2}
\end{eqnarray}
The discussion below is conducted for the energy, i.e. for a scalar operator $H$ belonging to the trivial irreducible representation $\lambda=0$ characterized by $S^{0}_{ba}(g)=\delta_{ab}$. However, such a discussion can be extended to any irreducible tensor operator~\cite{sadoudi10a}.

For nuclear structure, two groups are of particular importance as discussed in the introduction, i.e. $SO(3)$ for rotations in the three-dimensional space and $U(1)$ for rotations in the gauge space associated with particle number. The group of spatial translations is also essential but corresponds to a symmetry that is strongly broken in all nuclei and that does not need to be exactly restored in practice. Consequently, Tab.~\ref{detailsgroup} gather useful elements that characterize $U(1)$ and $SO(3)$ such that the formulae given in the paper for a generic compact Lie group can be easily adapted to either of them.

\begin{table}[htbp]
\caption{\label{detailsgroup} Characteristics of $SO(3)$ and $U(1)$ relevant to the present study. The gauge angle of $U(1)$ is $\varphi \in [0,2\pi]$ whereas Euler angles parameterizing $SO(3)$ are $\Omega \equiv (\alpha,\beta,\gamma) \in [0,4\pi] \times [0,\pi] \times [0,2\pi]$. The one-dimensional irreducible representations of $U(1)$ are labeled by $m \in \mathbb{Z}$ whereas the $(2J\!+\!1)$-dimensional ones of $SO(3)$ are labeled by $2J \in \mathbb{N}$ and are given by the so-called Wigner functions ${\cal D}^{J}_{MM'}(\Omega)$, where $(2M,2M')\in
\mathbb{Z}^2$ with $-2J\leq 2M,2M' \leq+2J$. For $U(1)$, one has $l(N)=N^2$, whereas for $SO(3)$, with the choice $C_i \equiv J_z$, one has  $l(J)=\hbar^2 J(J+1)$ and  $g(M)=M\,\hbar$.}
\small\rm
\begin{tabular}{|c|ccccccccc|}
\hline
${\cal G}$ &  $g$ &  $dm(g)$ & $v_{{\cal G}}$ & $\{C\}$ & $\Lambda$ & $C_i$ & $R(g)$ & $S^{\lambda}_{ab}(g)$ & $d_{\lambda}$ \\
\hline
$U(1)$ &  $\varphi$  & $d\varphi$  & $2\pi$ & $N$ & $N^2$ & - & $e^{iN\varphi}$& $e^{im\varphi}$ & $1$ \\
$SO(3)$ &  $\alpha,\beta,\gamma$  &  $\sin \beta d\alpha d\beta d\gamma$ & $16\pi^2$ & $\vec{J}$ & $J^{2}$ & $J_z$ & $e^{-i\alpha \hat{J}_{z}} \, e^{-i\beta
\hat{J}_{y}} \, e^{-i\gamma \hat{J}_{z}}$ &  ${\cal D}^{J}_{MM'}(\Omega)$ & $2J+1$ \\
\hline
\end{tabular}
\end{table}

\section{Wave-function-based method}
\label{wfbasedmethod}

The present section describes what we denote as a wave-function-based method where energy kernels are {\it explicitly and strictly} computed as matrix elements of a Hamilton operator that does {\it not} depend on the wave-function it is used with, e.g. $H$ is not an effective operator depending on the density of the system.

\subsection{Symmetry breaking}

In the present discussion, states $| \Theta^{\lambda a} \rangle$ defined above can be eigenstates of $H$, which corresponds to handling exact solutions of the many-body problem. In general though, we will only consider states that are approximations to those exact eigenstates of $H$. According to our introductory discussion, a normalized, symmetry-breaking state $| \Theta \rangle$ takes the form
\begin{eqnarray}
| \Theta \rangle  &=& \sum_{\lambda a}  c_{\lambda a}  | \Theta^{\lambda a} \rangle \,\,\, , \label{expanSSBstate}
\end{eqnarray}
where $\sum_{\lambda a}  |c_{\lambda a}|^2=1$. Using either Eq.~\ref{eq:ten:def1} or Eqs.~\ref{eq:ten:def2} and~\ref{expanSSBstate}, one can easily prove that the average energy
\begin{equation}
E \equiv \frac{\langle  \Theta | H | \Theta \rangle}{\langle  \Theta | \Theta \rangle}   \,\,\, , \label{SSBenergy}
\end{equation}
is a scalar under all transformations of ${\cal G}$. However, such an energy cannot be labeled by good quantum numbers $(\lambda, a)$, which is the fingerprint of the symmetry-breaking character of the many-body state $| \Theta \rangle$.

\subsection{Symmetry restoration}
\label{symrestorationWF}

One extends the definition of the symmetry breaking energy $E$ (Eq.~\ref{SSBenergy}) to the so-called energy {\it kernel} $E[g',g]$ through
\begin{eqnarray}
E[g',g] &\equiv& \frac{\langle  \Theta | R^{-1}(g') H R(g)  | \Theta \rangle}{\langle  \Theta | R^{-1}(g') R(g) | \Theta \rangle} \,\,\, , \label{SSBenergykernel0}
\end{eqnarray}
where the norm overlap kernel is $N[g',g] \equiv \langle  \Theta | R^{-1}(g') R(g) | \Theta \rangle$. Expanding $| \Theta \rangle$ according to Eq.~\ref{expanSSBstate} and using Eq.~\ref{unitarity}, one obtains
\begin{eqnarray}
E[g',g] \, N[g',g] &=&  \sum_{\lambda ab} c^{\ast}_{\lambda a} \, c_{\lambda b}\, E^{\lambda} \, S^{\lambda}_{ab}(g\!-\!g') \,\,\, , \label{SSBnormkernel0} \\ \label{SSBenergykernel}
N[g',g] &=&  \sum_{\lambda ab} c^{\ast}_{\lambda a} \, c_{\lambda b} \, S^{\lambda}_{ab}(g\!-\!g') \,\, \, , \label{SSBnormkernel}
\end{eqnarray}
where the symmetry-restored energies $\langle  \Theta^{\lambda a} | H  | \Theta^{\lambda' a'} \rangle \equiv E^{\lambda} \, \delta_{\lambda\lambda'} \, \delta_{aa'}$ provide the eigen-spectrum of $H$ if one is working with its exact eigenstates. Expressions~\ref{SSBenergykernel0} and~\ref{SSBnormkernel} correspond to the double expansion over the volume of the group
\begin{equation}
F[g',g] \equiv  \sum_{\lambda\lambda'} \sum_{aba'b'} F^{\lambda\lambda'}_{aba'b'} \, S^{\lambda}_{ab}(g') \, S^{\lambda}_{a'b'}(g)  \,\,\, , \label{genericexpansion}
\end{equation}
applied to functions $F[g',g]=f[g-g']$ that in fact only depend on the difference of the two arguments. In such a case, the double expansion reduces to a single expansion whose coefficients $f^{\lambda}_{ab}$ are $e^{\lambda}_{ab}\equiv c^{\ast}_{\lambda a} \, c_{\lambda b}\, E^{\lambda}$ and $n^{\lambda}_{ab} \equiv c^{\ast}_{\lambda a} \, c_{\lambda b}$ for the functions of interest $e[g\!-\!g']\,n[g\!-\!g']$ and $n[g\!-\!g']$, respectively. Given that coefficients $f^{\lambda}_{ab}$ transform specifically under any member of ${\cal G}$\footnote{The corresponding law is easily obtained from the transformation of $S^{\lambda}_{ab}(g)$.}, the ratio of two such objects, e.g. $E^{\lambda}$, transforms as a scalar.

Starting from $E[g',g]$ and $N[g',g]$, and given the orthogonality relationship
\begin{equation}
\int_{{\cal G}} dm(g) \, S^{\lambda \, \ast}_{ab}(g)  \, S^{\lambda'}_{a'b'}(g) = \frac{v_{{\cal G}}}{d_\lambda} \, \delta_{\lambda\lambda'} \, \delta_{aa'} \, \delta_{bb'} \,  \,\,\, , \label{orthogonality}
\end{equation}
one can perform the integration
\begin{equation}
\left(\frac{d_\lambda}{v_{{\cal G}}}\right)^2 \int \!\! \int_{{\cal G}}\! dm(g')\, dm(g) \, S^{\lambda}_{ca}(g') \,  S^{\lambda \, \ast}_{cb}(g) \, E[g',g] \, N[g',g] = e^{\lambda}_{ab} \,\,\, , \label{projectedenergy}
\end{equation}
to extract the energy $E^{\lambda}$ associated with states $| \Theta^{\lambda a} \rangle$ spanning the irreducible representation $\lambda$\footnote{The fact that $E[g',g]$ and $N[g',g]$ only depend on the difference $g\!-\!g'$ can be exploited to extract $e^{\lambda}_{ab}$ through a single integral rather than through a double integral as in Eq.~\ref{projectedenergy}. The reason why we keep explicitly two integrals here will only become clear in the last section of the paper (see Sec.~\ref{newconstraintsexact}).}. Such a symmetry restoration stage is denoted as a {\it multi-reference} method in the sense that, while the energy computed through Eq.~\ref{SSBenergy} involves a single reference state $| \Theta \rangle$, the extraction of $E^{\lambda}$ involves the set of references states $| \Theta (g) \rangle \equiv R(g) | \Theta \rangle$ obtained from $| \Theta \rangle$ through all transformations of ${\cal G}$.

\subsection{Transfer operator}
\label{tradition}

The above presentation of the symmetry restoration will be particularly suited to making the connection with the MR-EDF approach discussed in Sec.~\ref{EDFmethod}. Still, it is worth noting that Eq.~\ref{projectedenergy} is usually obtained within the wave-function-based approach by first introducing the transfer operator~\cite{ring80a}
\begin{equation}
P^{\lambda}_{ab} \equiv  \frac{d_{\lambda}}{v_{{\cal G}}}  \int_{{\cal G}} dm(g) \, S^{\lambda \, \ast}_{ab}(g) \, R(g) \,\,\, , \label{transferoperator}
\end{equation}
to extract the state with good symmetries
\begin{equation}
| \Theta^{\lambda a} \rangle = \frac{1}{c_{\lambda b}} \, P^{\lambda}_{ab} | \Theta \rangle \,\,\, , \label{projectedstate}
\end{equation}
from which Eq.~\ref{projectedenergy} can be recovered by taking the average value of $H$.

\subsection{Particular case of interest}
\label{particularcase}

Let us consider the particular case where the symmetry-breaking state $| \Theta \rangle$ is taken as a {\it product} state $| \Phi \rangle$ of the Bogoliubov type. In this case, the method at play corresponds to symmetry-unrestricted and symmetry-projected Hartree-Fock-Bogoliubov approximations~\cite{ring80a}. In such a situation, one can use the standard Wick theorem~\cite{wick50a} to express the symmetry-breaking energy $E$ (Eq.~\ref{SSBenergy}) as a specific functional $E[\rho,\kappa,\kappa^{\ast}]$ of the one-body density matrices computed from the symmetry breaking state and given, in an arbitrary single-particle basis, by
\begin{equation}
\label{rho}
\rho_{ij} \equiv \frac{\langle \Phi | a^{\dagger}_{j} a_{i}| \Phi \rangle}
                {\langle \Phi  | \Phi  \rangle} \,\, , \,\,
\kappa_{ij} \equiv \frac{\langle \Phi | a_{j} a_{i}| \Phi \rangle}
                {\langle \Phi | \Phi \rangle} \,\, , \,\,
\kappa^{\ast}_{ij} \equiv \frac{\langle \Phi | a^{\dagger}_{i} a^{\dagger}_{j}| \Phi \rangle}
                {\langle \Phi | \Phi \rangle} \, .
\end{equation}
The explicit form of $E[\rho,\kappa,\kappa^{\ast}]$ depends on the specific Hamiltonian $H$, e.g. on whether it is local or non local in space and on whether it contains a three-body force, a four-body force\ldots

As for the symmetry restoration stage, the energy kernel at play (Eq.~\ref{SSBenergykernel}) can be computed in this case using the {\it generalized} Wick theorem~\cite{balian69a} such that $E[g',g]=E[\rho^{g'g},\kappa^{g'g},\kappa^{gg'\,\ast}]$, i.e. the {\it transition} energy kernel is expressed through the same functional as the symmetry-breaking energy except that diagonal symmetry-breaking one-body density matrices are replaced by {\it transition} one-body density matrices involving the two transformed product states $| \Phi (g) \rangle$ and $| \Phi (g') \rangle$, i.e.
\begin{equation}
\label{transrho}
\rho^{g'g}_{ij} \equiv \frac{\langle \Phi (g') | a^{\dagger}_{j} a_{i}  | \Phi (g) \rangle}
                {\langle \Phi (g')  | \Phi (g) \rangle} \,\, , \,\,
\kappa^{g'g}_{ij} \equiv \frac{\langle \Phi (g') | a_{j} a_{i} | \Phi (g) \rangle}
                {\langle \Phi (g') | \Phi (g) \rangle} \,\, , \,\,
\kappa^{gg' \, \ast}_{ij} \equiv \frac{\langle \Phi (g') | a^{\dagger}_{i} a^{\dagger}_{j} | \Phi (g) \rangle}
                {\langle \Phi (g') | \Phi (g) \rangle} \, .
\end{equation}

\subsection{Actual implementation}
\label{practical}

In practice, states $| \Theta^{\lambda a} \rangle$ only provide approximations to exact eigenstates. This is for instance the case when starting from a symmetry-breaking product state as discussed in Sec.~\ref{particularcase}. When dealing with a non abelian group, one must actually consider an arbitrary linear combination of states spanning a given irreducible representation such that mixing coefficients are determined through the minimization of the resulting energy. This corresponds to considering that the link between the symmetry-restored states of interest and the symmetry-breaking one is in fact given by
\begin{equation}
| \Theta^{\lambda a} \rangle \equiv \sum_b g^{\lambda b} \, P^{\lambda}_{ab} | \Theta \rangle \,\,\, , \label{projectedstate2}
\end{equation}
rather than by Eq.~\ref{projectedstate}, and to determining the $\{g^{\lambda b}\}$ through the minimization of
\begin{equation}
E^{\lambda} \equiv \frac{\langle \Theta^{\lambda a} | H | \Theta^{\lambda a} \rangle}{\langle \Theta^{\lambda a} | \Theta^{\lambda a} \rangle} = \frac{\sum_{bb'} g^{\lambda b \ast} \, g^{\lambda b'} \, e^{\lambda}_{bb'}}{\sum_{bb'} g^{\lambda b \ast} \, g^{\lambda b'} \, n^{\lambda}_{bb'}} \,\,\, , \label{projectedenergy2}
\end{equation}
with $e^{\lambda}_{bb'}$ defined by Eq.~\ref{projectedenergy} and $n^{\lambda}_{bb'}$ given by a similar equation for the kernel $N[g',g]$. Such a minimization leads to solving a Hill-Wheeler-Griffin equation~\cite{hill53a,griffin57a}.

\section{EDF-based method}
\label{EDFmethod}

We are now interested in the nuclear EDF formalism within which energy kernels are formulated as functionals of the one-body (transition) densities defined by Eqs.~\ref{rho} and~\ref{transrho}. Although extensions can be envisioned, the standard EDF method is based on the use of simple symmetry-breaking product states of the Bogoliubov type. This choice stems from the historical construction of the nuclear EDF method by analogy with symmetry-unrestricted and symmetry-projected Hartree-Fock-Bogoliubov approximations (Sec.~\ref{particularcase})~\cite{ring80a}.

\subsection{SR-EDF step}
\label{SREDFmethod}

The SR-EDF method~\cite{bender03a} relies on computing the analog to the symmetry-breaking average energy $E$ (Eq.~\ref{SSBenergy}) as an a priori {\it general} functional ${\cal E}[\rho,\kappa,\kappa^{\ast}]$. As opposed to what was considered in Sec.~\ref{particularcase}, the symmetry-breaking energy ${\cal E}[\rho,\kappa,\kappa^{\ast}]$ is {\it not} computed from the average value of a genuine operator $H$. As a result, specific constraints must be imposed onto the functional form of ${\cal E}[\rho,\kappa,\kappa^{\ast}]$ for it to be a scalar under all transformations of ${\cal G}$; i.e. under transforming $| \Phi \rangle$ and the densities $\rho$, $\kappa$, $\kappa^{\ast}$ constructed from it. For a quasi-local functional of the Skyrme type, we refer the reader to Refs.~\cite{dobaczewski96a,perlinska04a,rohozinski10a} for the formulation of such constraints. For actual parameterizations of the nuclear EDF, we refer the reader to Ref.~\cite{bender03a}.

\subsection{MR-EDF step}
\label{MREDFmethod}

The MR-EDF step is built from the SR-EDF by analogy to the wave-function-based method formulated in terms of product states (Sec.~\ref{particularcase}). As a result, the analog of the energy kernel $E[g',g]$ (Eq.~\ref{SSBenergykernel}) is naturally introduced over the volume of ${\cal G}$ through the definition ${\cal E}[g',g]\equiv{\cal E}[\rho^{g'g},\kappa^{g'g},\kappa^{gg'\,\ast}]$, where ${\cal E}[\rho,\kappa,\kappa^{\ast}]$ is the SR-EDF kernel. From ${\cal E}[g',g]$, one extracts
\begin{equation}
\varepsilon^{\lambda}_{ab} \equiv \left(\frac{d_\lambda}{v_{{\cal G}}}\right)^2 \int_{{\cal G}} \! \int_{{\cal G}}\! dm(g')\, dm(g) \, S^{\lambda}_{ca}(g') \,  S^{\lambda \, \ast}_{cb}(g) \, {\cal E}[g',g] \, N[g',g] \,\,\, , \label{MREDFenergy}
\end{equation}
by analogy to Eq.~\ref{projectedenergy}. Whereas in the wave-function-based method one could explicitly demonstrate the identity $e^{\lambda}_{ab} = c^{\ast}_{\lambda a} \, c_{\lambda b} \, E^{\lambda}$, this is not the case in the EDF approach where there no possibility in general to perform the equivalent to the derivation that led to Eq.~\ref{SSBenergykernel0}. Equation~\ref{MREDFenergy} corresponds simply to the application of expansion~\ref{genericexpansion} to the {\it function} ${\cal E}[g',g]$ over the irreducible representations of the group, without any reference to many-body states with good quantum numbers. As a matter of fact, and contrarily to what is often stated~\cite{bender03a}, $\varepsilon^{\lambda}_{ab}$ is {\it not} computed from a projected state in the MR-EDF method, i.e. the transfer operator $P^{\lambda}_{ab}$ {\it cannot} be factorized explicitly in Eq.~\ref{MREDFenergy}. However, one can {\it implicitly}  relate the MR-EDF energy ${\cal E}^{\lambda}$ to the projected state $| \Phi^{\lambda a} \rangle$ obtained from $| \Phi \rangle$ as in Eq.~\ref{projectedstate2}. With this in mind, it is natural and customary~\cite{bender03a,lacroix09a} to {\it define} the symmetry-restored energy ${\cal E}^{\lambda}$ from $\varepsilon^{\lambda}_{ab}$ through the analog of Eq.~\ref{projectedenergy2}, i.e.
\begin{equation}
{\cal E}^{\lambda} \equiv \sum_{bb'} g^{\lambda b \ast} \, g^{\lambda b'} \, \varepsilon^{\lambda}_{bb'}/\sum_{bb'} g^{\lambda b \ast} \, g^{\lambda b'} \, n^{\lambda}_{bb'} \,\,\, , \label{MREDFenergy2}
\end{equation}
where the $\{g^{\lambda b}\}$ are determined through the minimization of ${\cal E}^{\lambda}$.

\subsection{Puzzling questions}
\label{puzzling}

We have clarified in previous sections that SR- and MR-EDF methods have been empirically constructed by analogy to symmetry-unrestricted and symmetry-projected Hartree-Fock-Bogoliubov approximations. The key difference with the latter is that the energy kernels at play in the EDF method are {\it not} defined as matrix elements of a genuine operator between wave-functions. For the rest, expressions utilized in both approaches, in particular regarding the extraction of the symmetry-restored energy (Eqs.~\ref{projectedenergy} and~\ref{projectedenergy2} versus Eqs.~\ref{MREDFenergy} and~\ref{MREDFenergy2}), look totally alike. Still, puzzling questions remain to be raised.

As mentioned in Sec.~\ref{SREDFmethod}, one must require at the symmetry-breaking level that the SR energy ${\cal E}[\rho,\kappa,\kappa^{\ast}]$ is a scalar under all transformations of ${\cal G}$. Such a requirement has led to formulating a set of constraints on the functional form of ${\cal E}[\rho,\kappa,\kappa^{\ast}]$~\cite{dobaczewski96a,perlinska04a,rohozinski10a}. The next question one may ask is the following: are those constraints imposed on the energy kernel ${\cal E}[\rho,\kappa,\kappa^{\ast}]$ at the SR level sufficient to making the MR-EDF method described in Sec.~\ref{MREDFmethod} well defined, in particular from a symmetry standpoint? In particular, one may wonder whether the fact that the energy kernel ${\cal E}[g',g]$, which is the key ingredient to the MR-EDF calculation, is not computed as the matrix element of a (genuine) operator makes the method ill-defined in any way?

As a matter of fact, a set of physical constraints to be imposed on ${\cal E}[g',g]$ have already been worked out~\cite{robledo07a}. The facts (i) that the MR energy should be real, (ii) that the SR-EDF should be recovered from the Kamlah expansion and (iii) that the Random Phase Approximation based on the SR-EDF ${\cal E}[\rho,\kappa,\kappa^{\ast}]$ should be recovered as a limit of the MR-EDF calculation~\cite{jancovici64a}, has helped limiting the energy kernel ${\cal E}[g',g]$ to depend on transition densities only, e.g. ${\cal E}[g',g]\equiv{\cal E}[\rho^{g'g},\kappa^{g'g},\kappa^{gg'\,\ast}]$.

The aim of the present contribution is to elaborate further on the question raised above and to discuss a path that could be followed to constrain more tightly MR-EDF calculations. References~\cite{doba07a,lacroix09a,bender09a,duguet09a} have already provided important elements in the case of $U(1)$, i.e. for particle-number restoration (PNR). Let us recall the main outcome of those studies prior to formulating the problem to be addressed.

\subsection{Lessons learnt from particle-number restoration}

Equation~\ref{SSBenergykernel} applied to $U(1)$ provides the Fourier decomposition $\varepsilon[\varphi] \, n[\varphi] = \sum_{N \in \mathbb{Z}} c^2_N \, {\cal E}^{N} \, e^{iN\varphi}$ of the periodic function $\varepsilon[\varphi] \, n[\varphi]$ over $[0,2\pi]$. From a mathematical standpoint, the sum runs a priori over all irreducible representations of the group, i.e. over both positive and negative integers $N$. From a physics point of view though, the label $N$ denotes the particle number of the physical system. Consequently, the sum should actually only run over positive integers, i.e. one should find $c^2_N {\cal E}^{N}=0$ and ${\cal E}^{N}=0$ for $N \leq 0$. In the wave-function-based method, such a result is indeed obtained from the fact that $E^N$ is computed as the average value of $H$ in $| \Phi^{N} \rangle$, the latter being zero~\cite{bender09a} for $N \leq 0$. In the EDF context, however, it was demonstrated~\cite{bender09a,duguet09a} that Fourier components ${\cal E}^{N}$ are a priori different from zero for $N \leq 0$, i.e. one usually obtains a non-zero symmetry-restored energy for negative particle numbers! This problem was shown~\cite{bender09a} to be related to unphysical mathematical properties of ${\cal E}[\varphi]$. Applying the regularization method proposed in Ref.~\cite{lacroix09a}, the cancelation of non-physical Fourier components was recovered~\cite{bender09a}. At the same time, components ${\cal E}^{N}$ for $N > 0$ were modified by up to 1 MeV, which is of the same order as the root-mean-square error on mass residuals reached by the best available particle-number-restored EDF mass fits~\cite{samyn04a}. This demonstrates the practical need of constraining further MR-EDF calculations in order to produce fully reliable results.

\section{Towards new constraints?}
\label{newconstraints}

The example discussed above is particularly enlightening given that clear-cut physical arguments can be used to argue that certain coefficients in the Fourier expansion $\varepsilon[\varphi] \, n[\varphi]$ should be strictly zero, although they are not if one does not pay particular attention to it. Recovering such physical features removes at the same time non-physical contaminations from other coefficients of the expansion~\cite{bender09a}. This proves that the MR-EDF method, as performed so far, faces the danger to be ill-defined and that new constraints on the energy kernel ${\cal E}[g',g]$ must be worked out in order to make the method physically sound. The regularization method proposed in Ref.~\cite{lacroix09a} that restores the validity of PNR can only be applied if the underlying EDF ${\cal E}[\rho,\kappa,\kappa^{\ast}]$ depends on integer powers of the density matrices~\cite{duguet09a}, which is an example of such a constraint.

For an arbitrary symmetry group, the situation might not be as transparent as for $U(1)$. Indeed, it is unlikely in general that certain coefficients of the expansion of ${\cal E}[g',g] N[g',g]$ over irreducible representations of the group are zero based on physical arguments. The challenge we face can be formulated in the following way: although expansion~\ref{genericexpansion} that underlines the MR-EDF method is sound from a group-theory point of view, mathematical properties deduced from a wave-function-based method must be worked out and imposed on ${\cal E}[g',g]$ to make $\varepsilon^{\lambda}_{ab}$ extracted from Eq.~\ref{MREDFenergy} physically sound. The rest of the present contribution is dedicated to briefly introducing an example of such a property in the case of $SO(3)$, i.e. for angular momentum restoration, that could be used to constrain ${\cal E}[\Omega',\Omega]$. Details of such an analysis will be reported elsewhere~\cite{sadoudi10a}.

\subsection{Mathematical property associated with angular-momentum conservation}
\label{newconstraintsexact}

We omit spin and isospin for simplicity and consider the rotationally-invariant nuclear Hamiltonian $H=T+V$ in which the central two-nucleon interaction
\begin{equation}
V \equiv \frac{1}{2} \int \!\!\int \! d\vec{r}_{1}  d\vec{r}_{2}  \, v(|\vec{r}_{1}-\vec{r}_{2}|) \, a^{\dagger}_{\vec{r}_{1}} \, a^{\dagger}_{\vec{r}_{2}} \, a_{\vec{r}_{2}} \, a_{\vec{r}_{1}} \,\,\, , \label{NNinteraction}
\end{equation}
is local, i.e. non-antisymmetrized matrix elements are defined as $\langle 1: \vec{r}_{1} ; 2: \vec{r}_{2} |V|  1: \vec{r}_{3} ; 2: \vec{r}_{4} \rangle \equiv v(|\vec{r}_{1}-\vec{r}_{2}|) \, \delta(\vec{r}_{1}-\vec{r}_{3}) \, \delta(\vec{r}_{2}-\vec{r}_{4})$, and in which three-nucleon and higher many-body forces are disregarded for simplicity. None of the conclusions drawn below would be modified by the inclusion of many-body forces or by using a non-local two-nucleon interaction. Operator $a^{\dagger}_{\vec{r}}$ ($a_{\vec{r}}$) creates (annihilates) a nucleon at position $\vec{r}$. Considering an eigenstate $| \Theta^{LM} \rangle$ of $\vec{L}^2$ and $L_z$, as well as using center of mass $\vec{R}\equiv(\vec{r}_1+\vec{r}_2)/2$ and relative coordinates $\vec{r}\equiv \vec{r}_1-\vec{r}_2$, the potential energy reads as
\begin{eqnarray}
V^{L} \equiv \frac{\langle \Theta^{LM} | V | \Theta^{LM} \rangle}{\langle \Theta^{LM} | \Theta^{LM} \rangle} &=& \frac{1}{2} \int \! d\vec{R}  \int \! d\vec{r} \,\, v(r) \, \rho^{[2] \, LMLM}_{\vec{R}\vec{r}}  \\
&\equiv&  \int \! d\vec{R}  \,\, V^{LM}(\vec{R}) \,\,\, ,  \label{exactpotentialenergy}
\end{eqnarray}
which defines a potential energy {\it density} $V^{LM}(\vec{R})$ in terms of the two-body density matrix $\rho^{[2] \, LMLM}_{\vec{R}\vec{r}} \equiv \langle \Theta^{LM} | \, a^{\dagger}_{\vec{r}_2} \, a^{\dagger}_{\vec{r}_1} \, a_{\vec{r}_1} \, a_{\vec{r}_2} \, | \Theta^{LM} \rangle/\langle \Theta^{LM} | \Theta^{LM} \rangle$. After tedious but straightforward calculations, one can demonstrate that~\cite{sadoudi10a}
\begin{eqnarray}
V^{LM}(\vec{R})  = \sum_{L'=0}^{2L} C^{LM}_{LML'0} \, \upsilon^{[2]}_{LL'}(R) \, Y^{0}_{L'}(\hat{R})  \,\,\, , \label{potentialenergydensity}
\end{eqnarray}
where the Clebsch-Gordan coefficient $C^{LM}_{LML'0}$ carries the dependence on $M$ while $Y^{m}_{l}$ denotes spherical harmonics. The weight $\upsilon^{[2]}_{LL'}(R)$ depends on the norm of $\vec{R}$ only and is related to a reduced matrix element of the two-body density matrix operator recoupled to a total angular momentum $L'$. The remarkable mathematical property identified through Eq.~\ref{potentialenergydensity} is that the scalar potential energy $V^L$ is obtained from an intermediate energy density $V^{LM}(\vec{R})$ whose dependence on the orientation of $\vec{R}$ is tightly constrained by the angular-momentum quantum number of the underlying many-body state $| \Theta^{LM} \rangle$, i.e. its expansion over spherical harmonics is limited to $L'\leq 2L$. Such a result is unchanged when adding the kinetic energy (density) to the potential energy (density) such that we restrict ourselves to the latter for simplicity. Of course, the energy eventually extracts the coefficient of the lowest harmonic, i.e. $V^L = \sqrt{4\pi} \int \! dR  \, \upsilon^{[2]}_{L0}(R)$.

\subsection{Wave-function-based symmetry-restoration method}
\label{newconstraintsWF}

Since property~\ref{potentialenergydensity} is general, it can also be obtained within the frame of the wave-function-based symmetry-restoration method presented in Sec.~\ref{wfbasedmethod}. Omitting again the kinetic energy for simplicity and using Eqs.~\ref{unitarity}-\ref{expanSSBstate}-\ref{SSBenergykernel}-\ref{NNinteraction}, the potential energy kernel reads
\begin{eqnarray}
V[\Omega',\Omega] \, N[\Omega',\Omega] &=& \frac{1}{2} \int
d\vec{R} \, d\vec{r} \; V(r) \, \langle \Theta \vert
R^{\dagger}(\Omega') \, \hat{\rho}^{[2]}_{\vec{R}\vec{r}} \,
R(\Omega) \vert \Theta \rangle \label{projenergydensity} \\ &=&
 \frac{1}{2}  \sum_{\{L,M\}} c^{\ast}_{L_1N_1} c_{L_2N_2} D^{L_1\dagger}_{N_1M_1}(\Omega')  D^{L_2}_{M_2N_2}(\Omega)  \int d\vec{R} \, d\vec{r} \; V(r)  \rho^{[2] \, L_1M_1 L_2M_2}_{\vec{R}\vec{r}} \,\,\,, \nonumber
\end{eqnarray}
where $\{L,M\}$ denotes a sum over the six angular-momentum quantum numbers appearing in the formula. Applying Eq.~\ref{projectedenergy} to the above expression (Eq.~\ref{projenergydensity}) provides, thanks to the orthogonality property~\ref{orthogonality}, the result
\begin{eqnarray}
V^{L} &=& \frac{(2L+1)^2}{(8\pi^2)^2 } \int d\Omega' d\Omega \, \frac{D^{L}_{KM}(\Omega')}{c^{ }_{LK}} \, \frac{D^{L \ast}_{KM}(\Omega)}{c^{\ast}_{LK} } \, V[\Omega',\Omega] \, N[\Omega',\Omega] \nonumber \\
 &=& \frac{1}{2} \int d\vec{R} \, d\vec{r} \; V(r) \,
 \rho^{[2]\, LM LM}_{\vec{R}\vec{r}} \,\,\,,
\end{eqnarray}
so that Eqs.~\ref{exactpotentialenergy} and~\ref{potentialenergydensity} are recovered. To obtain such a result it is mandatory to use the double-integral formulation of Eq.~\ref{projectedenergy} rather than the more standard single-integral formulation that takes advantage, from the outset, of the fact that $V[\Omega',\Omega]$ and $N[\Omega',\Omega]$ only depend on the difference $\Omega\!-\!\Omega'$. We thus insist on using the double-integral formulation in the present discussion.

\subsection{EDF-based symmetry-restoration method}
\label{newconstraintsEDF}

Let us now come back to the EDF method described in Sec.~\ref{EDFmethod}. The point is to underline the fact that property~\ref{potentialenergydensity} cannot be derived a priori given that the potential energy part of the kernel ${\cal E}[\Omega',\Omega]$ is {\it not} explicitly related to the two-body density matrix in this case. Taking a quasi-local Skyrme EDF as an example, although this can be easily adapted to non-local EDF of the Gogny type, the energy kernel takes the form
\begin{eqnarray}
{\cal E}[\Omega',\Omega] &=& {\cal E}[\rho^{\Omega'\Omega},\kappa^{\Omega'\Omega},\kappa^{\Omega\Omega'\,\ast}] \nonumber \\
&\equiv& \int d\vec{R} \,\,  {\cal E}(\rho^{\Omega'\Omega}(\vec{R}), \tau^{\Omega'\Omega}(\vec{R}), \vec{j}^{\Omega'\Omega}(\vec{R}),\ldots)  \,\,\, ,  \label{projEDFenergydensity}
\end{eqnarray}
where ${\cal
E}(\rho^{\Omega'\Omega}(\vec{R}),\tau^{\Omega'\Omega}(\vec{R}),\vec{j}^{\Omega\Omega'}(\vec{R}))$ is a general function of a set of {\it one-body} local transition densities built from the transition density matrices, e.g.
\begin{eqnarray}
\rho^{\Omega'\Omega}(\vec{R}) &\equiv& \sum_{ij} \varphi^{\ast}_{j} (\vec{R}) \, \varphi_{i} (\vec{R}) \, \rho^{\Omega'\Omega}_{ij} \,\,\, , \\
\tau^{\Omega'\Omega}(\vec{R})  &\equiv& \sum_{ij} \big[\vec{\nabla} \varphi^{\ast}_{j} (\vec{R})\big] \cdot \big[\vec{\nabla} \varphi_{i} (\vec{R})\big] \, \rho^{\Omega'\Omega}_{ij} \,\,\, , \\
\vec{j}^{\Omega'\Omega} (\vec{R}) &\equiv&  - \frac{i}{2} \sum_{ij} \Big\{   \varphi^{\ast}_{j} (\vec{R}) \, \big[\vec{\nabla} \varphi_{i} (\vec{R})\big] - \big[\vec{\nabla} \varphi^{\ast}_{j} (\vec{R})\big] \, \varphi_{i} (\vec{R}) \Big\} \, \rho^{\Omega'\Omega}_{ij} \,\,\, , \\
&\vdots& \nonumber
\end{eqnarray}
such that constraints imposed at the SR level~\cite{dobaczewski96a,perlinska04a,rohozinski10a} are fulfilled (see Secs.~\ref{SREDFmethod} and~\ref{puzzling}). Given such an EDF, there is no reason a priori that the energy density ${\cal E}^{LM}(\vec{R})$ extracted from Eqs.~\ref{MREDFenergy} and~\ref{MREDFenergy2} displays property~\ref{potentialenergydensity}; i.e. the angular dependence of ${\cal E}^{LM}(\vec{R})$ is likely to display harmonics $Y^{0}_{L'}(\hat{R})$ with $L'>2L$. One might argue that it is not an issue considering that the symmetry-restored energy ${\cal E}^L$ eventually relates to the harmonic $Y^{0}_{0}(\hat{R})$ only. However, a formalism that provides ${\cal E}^{LM}(\vec{R})$ with a spurious angular content will certainly also provide the coefficient ${\cal E}_{L0}(R)$ of the lowest harmonic with unphysical contributions. To state it differently, it is likely that constraining the MR-EDF kernels ${\cal E}[\rho^{\Omega'\Omega},\kappa^{\Omega'\Omega},\kappa^{\Omega\Omega'\,\ast}]$ to produce an energy density ${\cal E}^{LM}(\vec{R})$ that fulfils the mathematical property~\ref{potentialenergydensity} will impact at the same time the value of the weight ${\cal E}_{L0}(R)$, and thus the value of ${\cal E}^L$. To some extent, this is similar to the situation encountered with $U(1)$ where restoring the mathematical property that Fourier coefficients ${\cal E}^N$ with $N \leq 0$ should be strictly zero impacted the value of all other Fourier coefficients~\cite{bender09a}.

\section{Conclusions}

The present contribution reviews the notion of symmetry breaking and restoration within the frame of nuclear energy density functional (EDF) methods. Multi-reference (MR) EDF calculations are nowadays routinely applied with the aim of including long-range correlations associated with large-amplitude collective motions that are difficult to incorporate in a more traditional single-reference (SR), i.e.\ "mean-field", EDF formalism~\cite{bender03a}.

The framework for MR-EDF calculations was originally set-up by analogy with projection techniques and the Generator Coordinate Method (GCM), which are rigorously formulated only within a Hamiltonian/wave-function-based formalism~\cite{ring80a}. In the present work, we elaborate on key differences between wave-function- and energy-functional-based methods. In particular, we point to difficulties encountered to formulate symmetry restoration within the energy functional approach. The analysis performed in Ref.~\cite{bender09a} to tackle problems encountered in Refs.~\cite{almehed01a,anguiano01a,doba07a} for particle number restoration serves as a baseline. Reaching out to angular-momentum restoration, we identify in a wave-function-based framework a mathematical property of the energy density $E^{LM}(\vec{R})$ associated with angular momentum conservation that could be used to constrain EDF kernels. Consequently, possible future routes to better formulate symmetry restorations in EDF-based methods could encompass the following points.
\begin{enumerate}
\item The fingerprints left on the energy density $E^{LM}(\vec{R})$ by angular momentum conservation in a wave-function-based method could be exploited to constrain the functional form of the basic EDF ${\cal E}[\rho,\kappa,\kappa^{\ast}]$.
\item The regularization method proposed in Ref.~\cite{lacroix09a} to deal with specific spurious features of MR-EDF calculations should be investigated as to what impact it has on properties of the energy density ${\cal E}^{LM}(\vec{R})$ in the case of angular momentum restoration.
\item Similar mathematical properties extracted from a wave-function-based method could be worked out for other symmetry groups of interest and used to constrain EDF kernels.
\end{enumerate}
Efforts in those directions are currently being made~\cite{sadoudi10a}.

\ack

The authors wish to thank D. Lacroix for proofreading the manuscript and for useful discussions. T. D. wishes to thank M. Bender, P.-H. Heenen and D. Lacroix for long term collaborations on matters related to the subject of the present paper.

\end{document}